\documentclass[12pt]{iopart}
\usepackage{graphicx}% Include figure files
\usepackage{amssymb}
\begin{document}

\title{Evolution of dopant-induced helium nanoplasmas}

\author{S. R. Krishnan$^1$, Ch. Peltz$^2$,L. Fechner$^1$, V. Sharma$^1$, M. Kremer$^1$, B. Fischer$^1$, N. Camus$^1$, T. Pfeifer$^{1}$, J. Jha$^{3}$, M. Krishnamurthy$^{3}$, C. -D. Schr\"oter$^{1}$, J. Ullrich$^{1}$, F. Stienkemeier$^4$, R. Moshammer$^{1}$, Th. Fennel$^2$, M. Mudrich$^4$}

\address{$^1$Max-Planck-Institut f{\"u}r Kernphysik, 69117 Heidelberg, Germany\\
$^2$Institute of Physics, University of Rostock, Universit{\"a}tsplatz 3, D-18051 Rostock, Germany\\
$^3$Tata Institute of Fundamental Research, 1 Homi Bhabha road, Mumbai 400005, India\\
$^4$Physikalisches Institut, Universit\"at Freiburg, 79104 Freiburg, Germany}

\begin{abstract}
Two-component nanoplasmas generated by strong-field ionization of doped helium nanodroplets are studied in a pump-probe experiment using few-cycle laser pulses in combination with molecular dynamics simulations. High yields of helium ions and a pronounced, droplet size-dependent resonance structure in the pump-probe transients reveal the evolution of the dopant-induced helium nanoplasma. The pump-probe dynamics is interpreted in terms of strong inner ionization by the pump pulse and resonant heating by the probe pulse which controls the final charge states detected via the frustration of electron-ion recombination.
\end{abstract}


\section{Introduction}
Nanoplasmas generated by intense femtosecond laser pulses are being actively investigated for gaining insights into the ultrafast dynamics of highly excited  matter on the nanoscale, which features extraordinary characteristics. In particular the peculiar property of laser-driven nanoplasmas of emitting highly energetic particles and radiation opens up tremendous opportunities for applications as novel sources of radiation and for particle acceleration~\cite{sasi06,feme10}. Besides, the dynamics of multi-component nanoplasmas turns out to crucially impact the envisaged goal of realizing single-shot ultrafast diffraction imaging of large natural systems in the X-ray domain~\cite{Chapman}.

As a result of dedicated experiments as well as simulations, the behavior of single component nanoplasmas in intense near infrared fs laser pulses (10$^{14}$-$10^{16}\,$W/cm$^{2}$) is fairly well-understood (see~\cite{sasi06,feme10} and Refs. therein). These nanoplasmas are generated from neutral clusters of rare-gas or metal atoms via ionization by intense near-infrared (IR) laser pulses. The most interesting properties of rare-gas clusters in the IR domain result from their high energy absorption per atom once ionized, by far exceeding the values achievable in atomic jets or planar solid targets \cite{sasi06,feme10,dido96}. The generic picture of the underlying dynamics of laser-driven nanoplasmas is determined by the interrelation of the nanoplasma eigenfrequency $\omega_{res}$ and the frequency $\omega_{las}$ of the driving laser pulse. The dipolar eigenfrequency of the nanoplasma $\omega_{res}$ depends on the ionic charge density $\rho$, which for the spherical case is $\hbar\omega_{res}=\hbar\sqrt{e^2\rho/(3\varepsilon_0 m_e)}$~\cite{sasi06,feme10}. Resonant driving conditions ($\omega_{res}=\omega_{las}$) are achieved on sub- or few-picosecond timescales when the plasma is diluted by ionic expansion.

In spite of this general understanding of the plasma dynamics, experiments have revealed surprises in the case of nanoplasmas from two- or multi-component clusters: Intense IR pulse ionization of weakly doped (\textless 1\% doping) rare-gas clusters show enhanced electron and characteristic X-ray yields as compared their pristine counterparts under identical conditions \cite{jha2008collisionless,jhmk08}. While in these cases the location of the dopant atoms could not be determined, extensive studies on the intense IR field ionization of metal and rare gas clusters embedded at the center of He nanodroplets have been carried out in the group of K.-H. Meiwes-Broer \cite{feme10,dodi07,dofe06,fennel2007plasmon,kosc99}. These studies highlighted the role of the embedded dopant kernel in the nanoplasma dynamics and ionic motion thereof.

As compared to free metal clusters, resonance conditions were found to be reached at earlier delay times when measuring charge state spectra of the embedded species \cite{DoepEPJD2003,DoeppnerPCCP2007}. The appearance of He$^+$ and He$^{2+}$ ions was related to charge transfer collisions of highly charged metal kernel ions with He, in accord with the observed reduced maximum charging of the dopants. In large droplets, caging of ionization fragments was observed, which induces reaggregation and the formation of snowball complexes \cite{DoeppnerPCCP2007,DoeppnerJCP2007}.
The active role of the He droplet in the plasma dynamics was initially discerned by numerical simulations \cite{misa08,MikaberidzePRL2009,DoepPRL10,PeltzEPJD2011}. From molecular dynamics studies is was concluded that the He shell undergoes rapid inner ionization and selective resonant heating following the avalanche-like ionization of the dopant kernel~\cite{DoeppnerPCCP2007, PeltzEPJD2011}. As a consequence, well-separated multiple resonances as a function of pump-probe delay are predicted to occur in the absorption and photo electron spectra. However, no pronounced signature for such double resonance is seen in the pump-probe dependence of the final ion charge spectra of the dopant ions due to the complex interrelation of internal ionization and recombination dynamics.

In the present paper we report a combined experimental and numerical study which unambiguously demonstrates the dynamical evolution of the host He matrix. It extends our recent experiments using single 10\,fs pulses which have demonstrated the ignition of He droplets induced by only a few dopant atoms \cite{KrishnanPRL}.
Pump-probe measurements with 10\,fs pulses performed under the same experimental conditions together with the corresponding numerical simulations reveal a sensitive dependence of the optimal delays for high charging of He ion on the size of the nanodroplet and thus the explicit role played by the ionized He shell in the expansion dynamics of the composite nanoplasma. Furthermore, signatures for multiple resonances are predicted in the He ions yields from the doped nanodroplets. We show that it is crucial to account for the recombination dynamics to capture the complete evolution of this dopant-induced nanoplasma.

\section{Experiment}
\begin{figure}
\begin{center}{
\includegraphics[width=0.8\textwidth]{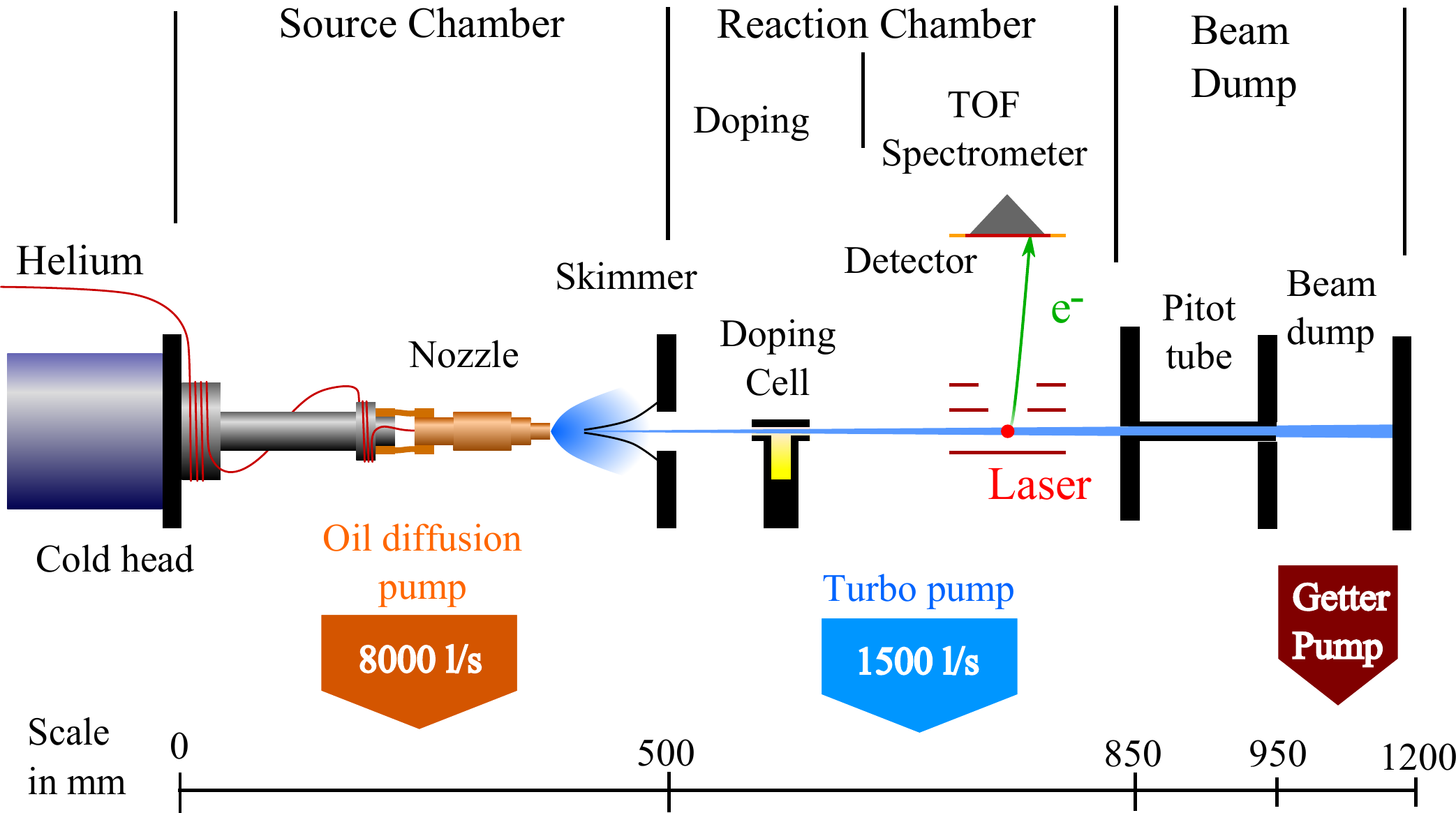}}
\caption{Experimental arrangement: Schematic
of the assembly consisting of the He nanodroplet source, the doping
chamber, the TOF spectrometer and the beam dump.}
\label{fig:Experimental-arrangement}
\end{center}
\end{figure}
$^{4}$He nanodroplets are an ideal host medium for designing well-defined
nanometer-sized two-component clusters \cite{ToenniesVilesovAnnRevPhysChem}.
Owing to their weak coupling to dopant atoms and to their superfluid
state, He droplets can pick up other rare gas atoms in a doping cell
that aggregate to form clusters immersed in the droplet interior \cite{ToenniesVilesovAnnRevPhysChem,FrankRev}. The experimental arrangement is schematically shown in Fig.~\ref{fig:Experimental-arrangement}.
A beam of He nanodroplets is produced by expanding
pressurized $^{4}$He gas (70 - 90 bar) through a nozzle $5\,\mu$m
in diameter maintained at a temperature of 15-25\,K. By varying the
nozzle temperature in this range the mean number of He atoms per droplet
is adjusted in the range 10$^{3}$-10$^{5}$. In a second vacuum chamber
further downstream, the skimmed droplet beam passes through a doping
cell consisting of a 3\,cm long cylindrical cell with two collinear
apertures (\O$\,=\,3\,$mm) to which a pressure gauge is directly attached
for monitoring the cell pressure. By leaking into the cell a controlled
amount of krypton (Kr) or xenon (Xe) rare gas using a dosing valve (leak rate $<10^{-10}$\,mbar\,l/s)
the mean number of dopants per nanodroplet $K$ can be controlled.
When taking into account the shrinkage of He droplets due to evaporation
of He atoms induced by the pick-up and cluster aggregation process,
$K$ can be determined from the cell pressure and the droplet size
according to a modified Poissonian pick-up statistics~\cite{kugo07}. This model is validated using Monte-Carlo simulations of the pick-up process as described in Ref.~\cite{bust11}.

Intense few-cycle laser pulses ($\sim10$\,fs) at a central wavelength
of 790\,nm with peak intensities in the range $10^{14}$ - $10^{15}$\,W/cm$^{2}$
are generated by a Ti-Sapphire based mode-locked laser system (Femtopower, Femtolasers GmbH, Vienna). Pairs of identical pulses are created using a Mach-Zehnder interferometer. The delay time between the first ``pump'' and the subsequent ``probe'' pulse is adjusted by varying the length of one of the arms of the interferometer by moving a pair of retro-reflecting mirrors mounted on a piezo-driven translation stage. The collinearly aligned pump and probe pulses are focused by a spherical mirror
(focal length $f=100\,$mm) into the beam of doped He nanodroplets. Photoions are detected by a time-of-flight
(TOF) spectrometer in the Wiley-McLaren geometry. The use of ultrashort pulses ensures that the laser couples purely to the electronic degrees of freedom of the (ionized) He nanodroplets whereas the ionic motion can be safely disregarded within the duration of the laser pulses.

\section{\label{sec:theory}Theory}
For modeling the intense laser-nanodroplet interaction we employ a classical molecular dynamics approach~\cite{FenPRL07b,PeltzEPJD2011}. Therein ions and plasma electrons are described  classically while atomic ionization events via tunnel and electron impact ionization are modeled quantum mechanically under the inclusion of local plasma field effects. Electron-ion recombination is taken into account for calculating ion charge state distributions. The key aspects relevant for the present study are sketched briefly below.

As model systems we consider He nanodroplets ($N_{He}$=5000, 10000 and 15000) doped with 20 Xe atoms. The nanodroplets are initialized as spheres (fcc structure) at the atomic density of liquid helium ($\rho=0.022$\,\AA$^{-3}$~\cite{ToenniesVilesovAnnRevPhysChem}). After inserting the Xe$_{20}$ core, He atoms with Xe neighbors closer than the equilibrium He-Xe distance of 4.15\,\AA\, are removed \cite{CheJCP73}. All atoms in the doped nanodroplets are initialized in their charge neutral state. Electron liberation into the nanodroplet environment (inner ionization) via tunnel ionization (TI) and electron impact ionization (EII) is described statistically via appropriate rates. The probability for TI is evaluated from the instantaneous Ammosov-Delone-Krainov rates~\cite{ADK_1986}, employing the local electric field. A successful TI event results in a new plasma electron at the classical tunnel exit. Electron impact ionization is evaluated from the Lotz cross sections \cite{LotzZP67}, taking into account local plasma field effects, such as depression of in-medium ionization potentials~\cite{FenPRL07b}. The resulting ions and electrons are then propagated classically in the laser field and under the influence of binary Coulomb interactions via
\begin{equation}
m_i\ddot{\bf r}_i=q_ie{\bf E}_{\rm las}-\nabla_{{\bf r}_i}\sum_{i\neq j}V_{ij},
\end{equation}
where $m_i$, $q_ie$, and ${\bf r}_i$ are the mass, charge, and position of the $i$-th particle and \mbox{${\bf E}_{\rm las}$} describes the laser electric field of two linearly polarized gaussian laser pulses. The pairwise Coulomb interaction $V_{ij}$ is described with a pseudopotential of the form
\begin{equation}
V_{ij}(r_{ij},q_1,q_2)=\frac{e^2}{4\pi \varepsilon_0}\frac{q_i q_j}{r_{ij}} \,{\rm erf}\left(\frac{r_{ij}}{s}\right)
\end{equation}
with the elementary charge $e$, the interparticle distance $r_{ij}$, their charge states $q_i$ and $q_j$, and a numerical smoothing parameter $s$. The latter regularizes the Coulomb interaction and offers a simple route to avoid classical recombination of electrons below the lowest possible quantum mechanical energy level. In our case, the smoothing parameter is determined by He and has a value of $s=0.67\,{\rm \AA}$. The time consuming evaluation of particle-particle interactions is accelerated using massive parallel computation techniques.

For the identification of resonance absorption in different regions of the doped droplet we perform a spatially resolved analysis of the energy absorption from the laser field. To this end the droplet is divided into time-dependent spherical regions $\Omega_{Xe}$ and $\Omega_{He}$ that contain the Xe core and the surrounding He droplet, respectively. The instantaneous power absorption in region $\Omega_k$ is determined from the dipole velocity by
\begin{equation}
P^{\Omega_k}(t)=\sum_{{\bf r}_i \in \Omega_k} e\,q_i \dot{\bf r}_i \cdot {\bf E}_{\rm las}(t),
\end{equation}
leading to the accumulated energy absorption
\begin{equation}
W_{\rm abs}^{\Omega_k}(t)=\int_{-\infty}^t P^{\Omega_k}(t') dt'.
\end{equation}

For a meaningful comparison of experimental and theoretical results, final ion charge spectra have to be determined from the simulation. Recent studies have shown that recombination has a crucial impact on the charge distributions and thus must be taken into account for predicting realistic charge spectra~\cite{DoepPRL10,PeltzEPJD2011,FenPRL07b}.
The two relevant mechanisms for recombination of quasi-free electrons with atomic ions after the laser excitation are radiative recombination and three-body recombination (TBR). As has been estimated previously~\cite{FenPRL07b}, radiative recombination can be neglected due to low rates~\cite{BetQM77}. TBR, i.\,e. electron capture after the collision of two quasi-free electrons in the vicinity of an ion, proceeds mainly to high Rydberg states of the ion and can therefore be treated classically. Hence, TBR is automatically included in reasonable approximation within the classical MD propagation. Besides, the dominant contribution of TBR, even higher order collisional recombination processes (four-body, five-body, etc.) are accounted for because of the fully microscopic description of classical particle-particle correlations. The key advantage of the direct microscopic treatment of recombination is that no approximations such as quasi charge-neutrality or a thermal electron velocity distribution need to be used. In addition, the local field effects due to screening and potentials of neighboring ions are included. To approximate the resulting charge spectra from the MD simulation, electrons are treated as recombined when bound to a specific ion after 1\,ps of propagation subsequent to the probe pulse. For details see~\cite{PeltzEPJD2011}.

\section{\label{sub:PPIntegral-ion-yields}Results}
\subsection{\label{sub:PP-dynamics}Pump-probe dynamics}
\begin{figure}
\begin{center}{
\includegraphics[width=0.5\textwidth]{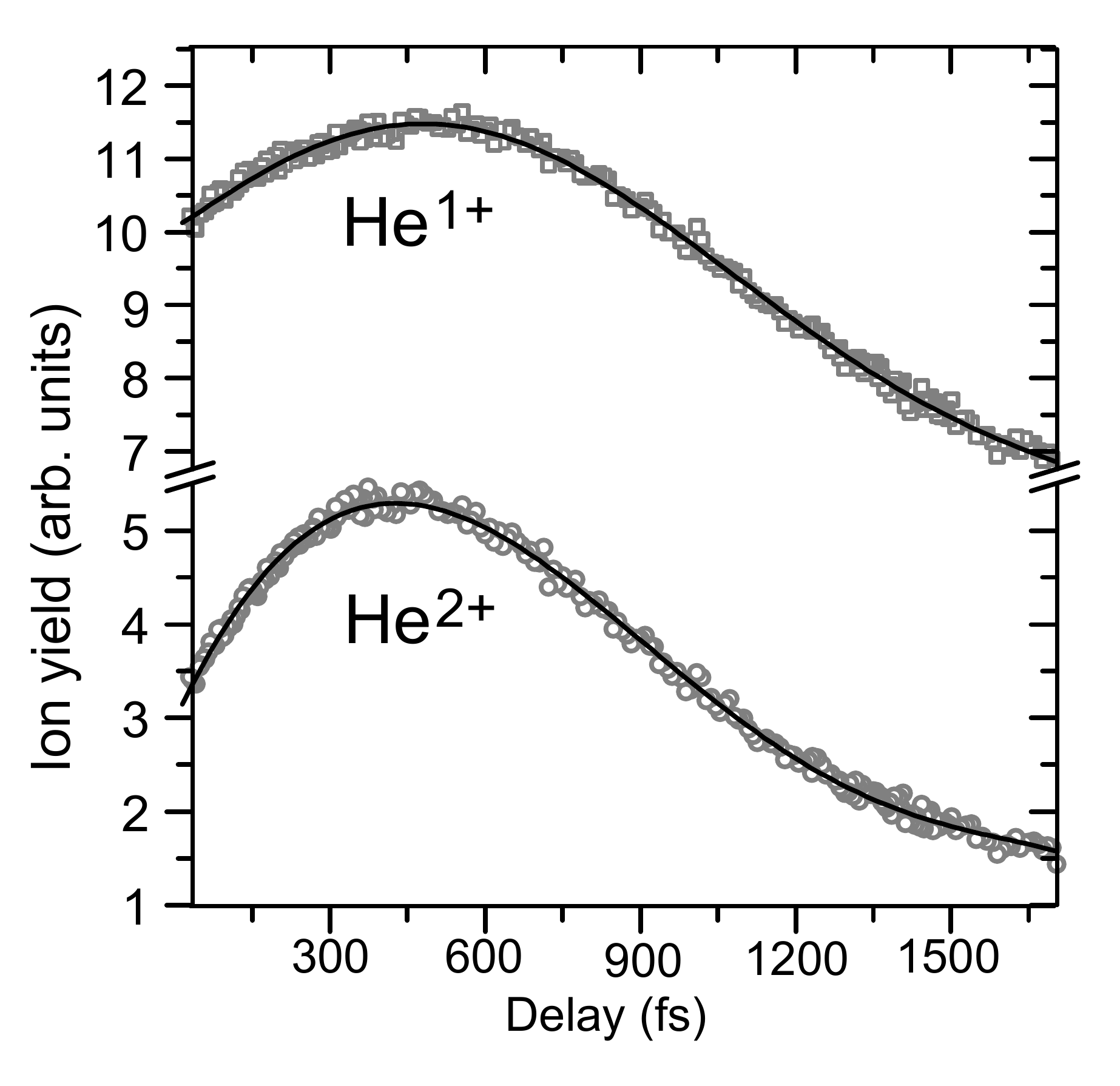}}
\caption{\label{fig:CompareHePandHe2P}Comparison of He$^{+}$ and He$^{2+}$
ion yields as a function of pump-probe delay when Xe$_{15}$@He$_{15000}$
nanodroplets are exposed to two identical pulses ($\sim10\,$fs) of
peak intensity $7\times10^{14}\,$W/cm$^{2}$. The He$^{2+}$
yield is maximized at a delay time of $457\,(\pm15)\,$fs and the
He$^{+}$ yield at $476\,(\pm15)\,$fs. Lines are to guide the eye.}
\end{center}
\end{figure}
Fig.~\ref{fig:CompareHePandHe2P} shows a typical example of the measured He$^{+}$ and He$^{2+}$ ion yields as a function of pump-probe delay. Both ion signals exhibit a pronounced pump-probe dynamics on the sub-ps time scale with maximum yields for $\tau_{opt}\approx0.45\,$ps delay.
Note that the He$^{2+}$ signal shows higher pump-probe contrast and reaches a maximum at a slightly smaller value of the optical delay ($\Delta \tau_{opt}\approx 20$fs).
%A similar trend was observed by D\"oppner et al.~\cite{dodi07}.

In a simplified picture, the existence of an optimal delay can be interpreted as a signature for resonantly enhanced charging~\cite{kosc99,DoeppnerPCCP2007,PeltzEPJD2011}. After the pump pulse ionizes the doped droplet and creates an overdense nanoplasma, the probe pulse excited the system resonantly after the appropriate degree of expansion. We recently reported a high degree of ionization in the He shell induced by a few dopant atoms at the center of the droplet~\cite{KrishnanPRL}. At the initial near-solid atomic densities prevalent within the droplet during the pump pulse, the dipolar eigenfrequency ($\hbar\omega_{res}\approx 3.2\,$eV) by far exceeds the frequency $\hbar\omega_{las}\approx 1.6\,$eV of the laser pulses at the wave length of 790\,nm. Hence, ionic expansion via Coulombic and hydrodynamic forces is required to reach resonant conditions. As a consequence, the absorption of energy from the probe laser pulse by the ionized droplet rises sharply, which is likely to enhance electron emission as well as electron-ion recombination taking place in the later stage of the expansion. A detailed analysis of the full pump-probe evolution of the nanoplasma is presented in the following section.

%{\bf TF to Marcel: everything in this para is correct, but I'm not sure if we really need this discussion- I leave it up to you} Most of the previous investigations of the expansion-induced nanoplasma resonance have been performed by varying pulse-widths. In
%practice, this is achieved by manipulating the gratings in the pulse compressor of a chirped-pulse amplification system to deliberately
%introduce a chirp in the pulses and thereby ''stretching'' them. While such studies reproduce the general time-dependent features of
%the expansion-induced resonance, care should be taken while interpreting and comparing the various results \cite{kumarappan2003electrons,jhmk08,zwdi99}.
%In these experiments often the pulse energy is kept constant while the pulse width is varied. Hence, the longer pulses have smaller peak
%intensities. Some threshold processes like ionization by barrier suppression~\cite{KrainovPhysUsp} are not triggered at smaller peak laser intensities.
%Sakabe et al. \cite{sakabe2006skinning} have shown that cluster ion kinetic energy spectra are indeed sensitive to atomic barrier
%suppression thresholds. Alternatively, the pulse energy can be varied along with the pulse width to keep the peak intensity constant. However,
%due to the limited pulse energy available from typical laser systems, maintaining a constant high intensity during a 10- or 50-fold variation
%of pulse widths is often not achievable in practice. Thus, two-pulse experiments are ideally suited for studying such dynamics.

%\subsection{\label{sub:PP-Droplet-size-dependence}Droplet size dependence}

\begin{figure}
\begin{center}{
\includegraphics[width=0.6\textwidth]{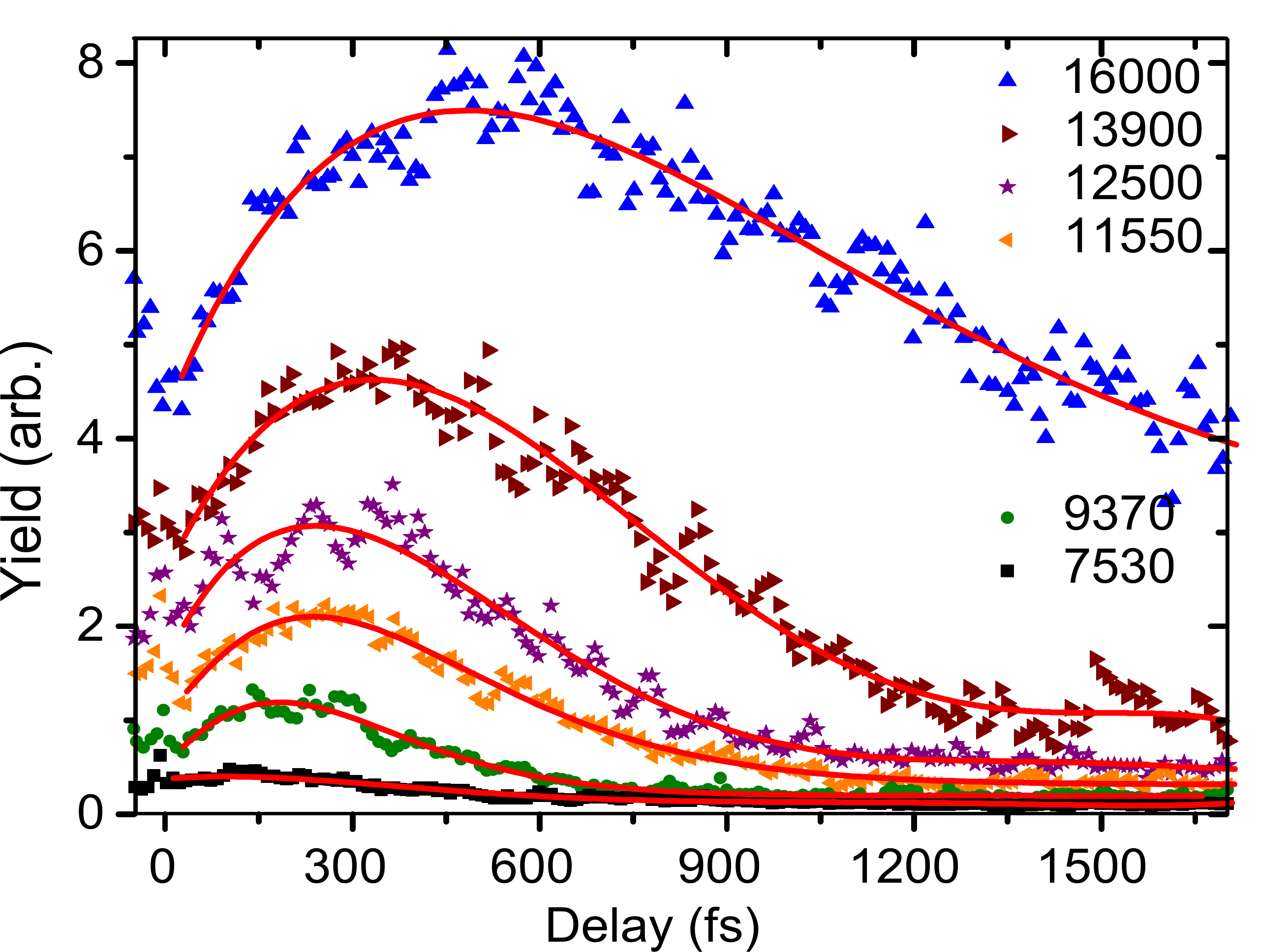}} % trim=10bp 560bp 300bp 840bp, clip=true
\caption{\label{fig:SizeDepPPHe2p}Delay dependence of He$^{2+}$ ion yields
for various droplet sizes (mean number of He atoms per droplet is
indicated in the legend) for the case of Xe doping. The peak intensity of
the pump and probe pulses is $7\times10^{14}\,$W/cm$^{2}$.
The mean number of doped atoms in the droplets is $15\pm3$ in the
case of Xe. The data points are fitted with a fifth-order polynomial
from which the optimal delay values ($\tau_{opt}$) are extracted.}
\end{center}
\end{figure}
\begin{figure}
\begin{center}{
\includegraphics[width=0.4\textwidth]{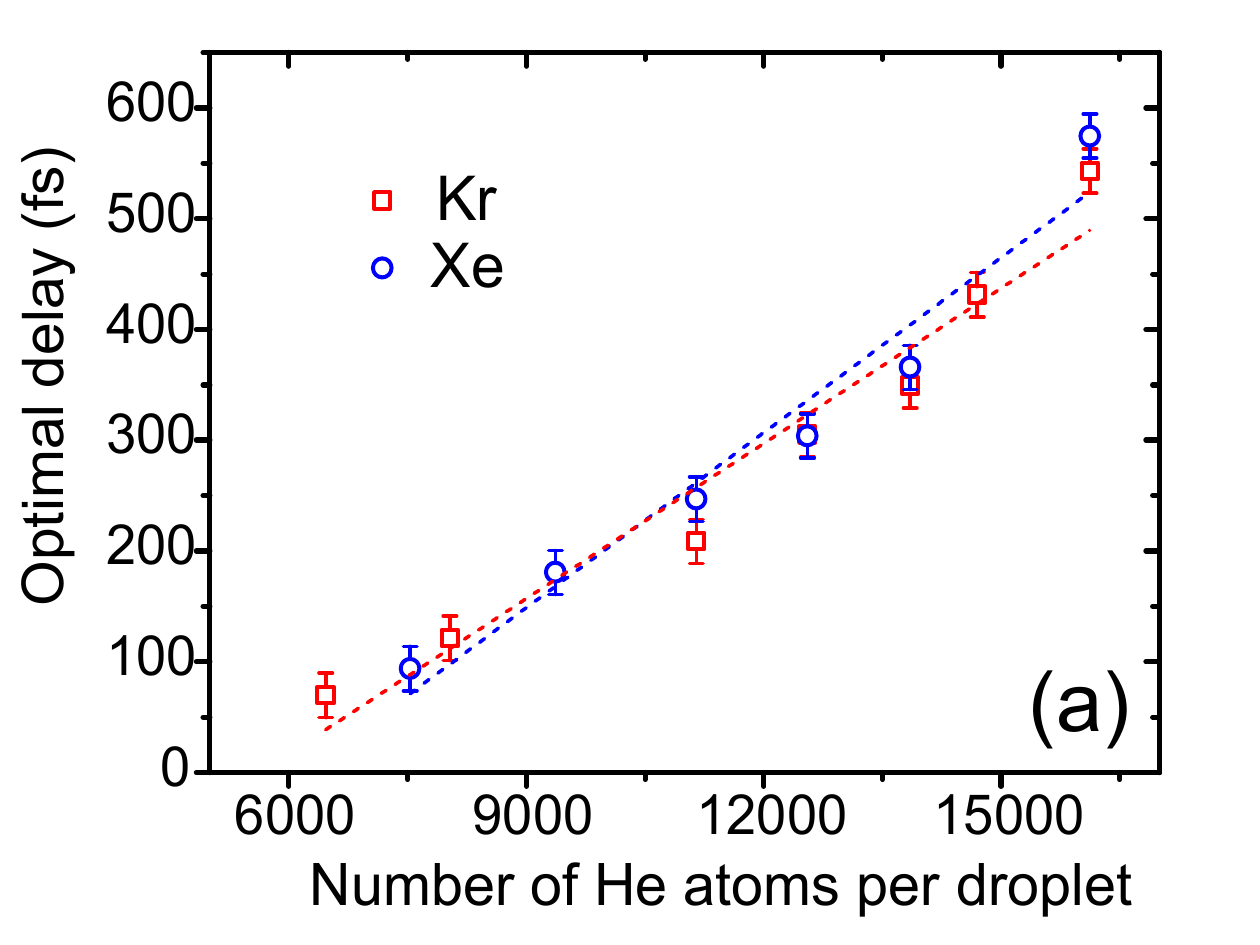}\includegraphics[width=0.4\textwidth]{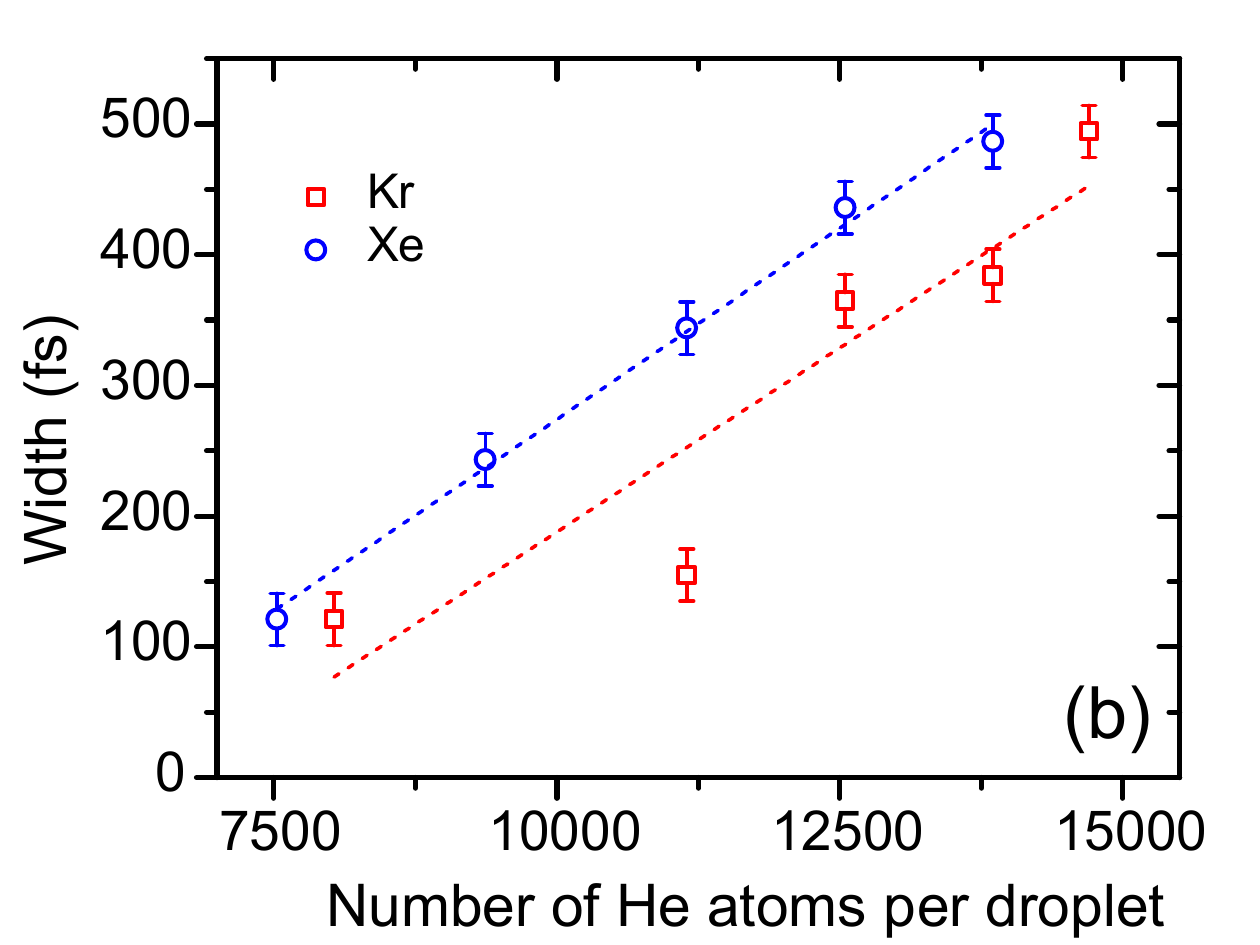}}
\caption{\label{fig:T-opt-XeKrAr}(a) Optimal delay times for Xe, Kr and Ar
doped He nanodroplets of various sizes. The Xe and Kr cases are the
same as shown in figure \ref{fig:SizeDepPPHe2p}. The corresponding
variation of the widths (FWHM) of ion yield curves
is shown in panel (b). The mean number of doped Xe and Kr atoms
is $15\,\pm3$ and $14\,\pm2.8$, respectively. The peak intensity
of the pump and probe pulses is $7\times10^{14}\,$W/cm$^{2}$.}
\end{center}
\end{figure}

The delay dependence of He ion yields are investigated for various experimental parameters including He droplet sizes for two different dopant species - Xe and Kr. The He$^{+}$ and He$^{2+}$ resonance curves peaked around 500\,fs turns out to be a robust feature with respect to variations of the laser pulse intensities as well as of the species and number of dopant atoms. The most pronounced peak shifting and broadening is observed when the size of the He droplets is varied. Fig.~\ref{fig:SizeDepPPHe2p} presents the delay dependence of He$^{2+}$ ion yields for the cases of Xe and Kr doping. The mean number of dopants per droplet ($\sim15$) is nearly the same in both cases. Clearly, the optimal delay $\tau_{opt}$ increases with increasing droplet sizes. This is similar to the delay dependence measurements of optical absorption by pure Xe clusters of different sizes observed by Zweiback et al. that were rationalized mainly by geometrical effects~\cite{zwdi99}. Figs.~\ref{fig:T-opt-XeKrAr}(a) and (b) illustrate the measured size dependences of the optimal delay and peak width (FWHM) as obtained from fitting the pump-probe data (Fig.~\ref{fig:SizeDepPPHe2p}) with a fifth-order polynomial. Both quantities are found to vary roughly linearly with the average number of He atoms per droplet in the considered size range.
% (goodness of fit better than 98 \%).

\subsection{\label{sub:theory}Simulation results}
In order to extract the mechanisms underlying the experimentally observed pump-probe dynamics we have performed MD simulations with doped He nanodroplets $\mathrm{Xe}_{20}\mathrm{He}_{N}$ exposed to two 10\,fs pulses at the intensity $I=7\times10^{14}\,{\rm W/cm}^2$. Note that this intensity is well below the threshold for barrier suppression ionization of He but sufficient for tunnel ionization of Xe~\cite{Augst}.

\begin{figure}[hbt]
\center
\includegraphics[width=0.9\columnwidth]{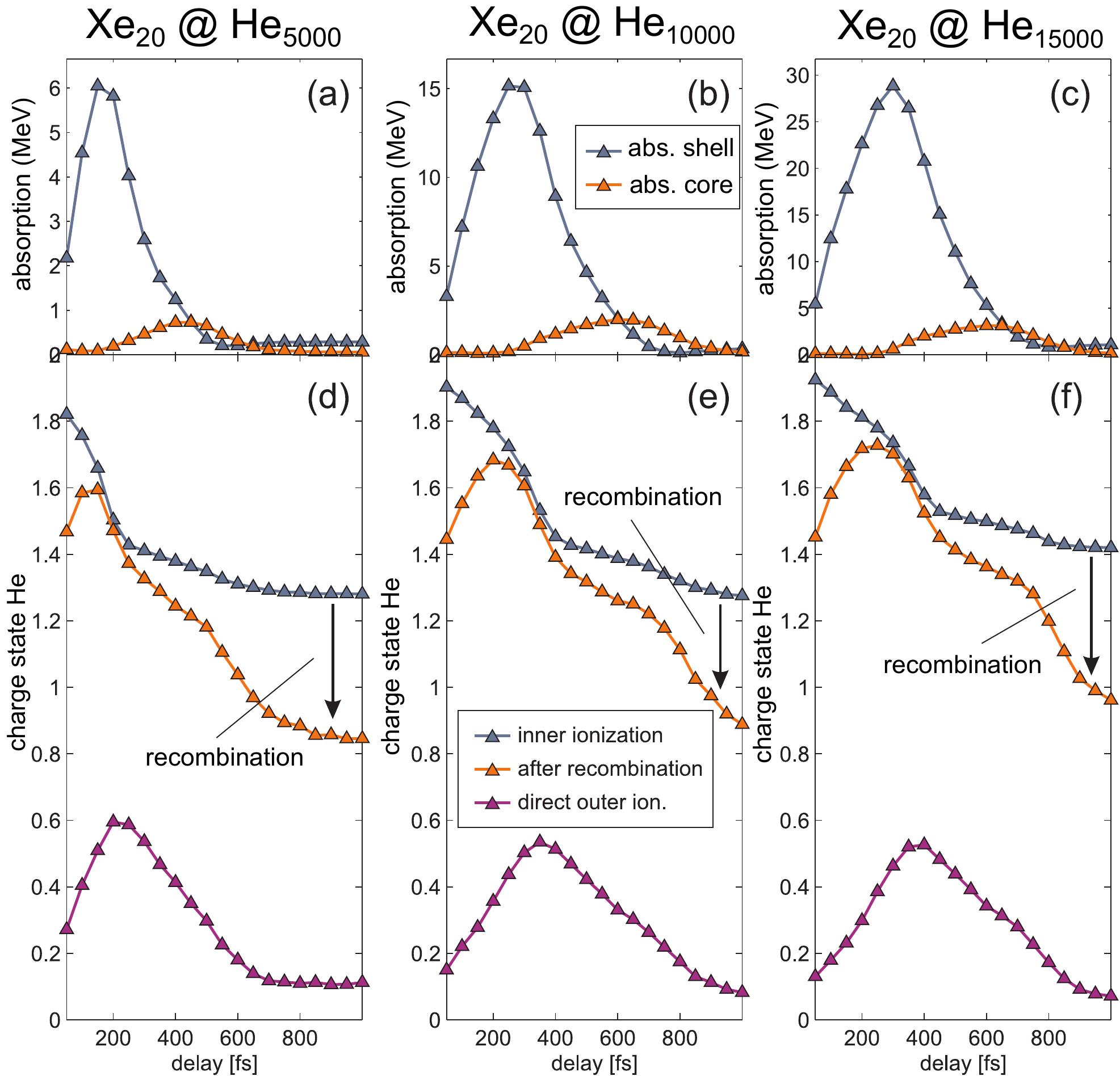} 
\caption{\label{fig:md_xehe} Calculated pump probe dynamics of $\mathrm{Xe}_{20}\mathrm{He}_{N}$ after exciation with two 10\,fs pulses (intensity: $I=7\times10^{14}\mathrm{W/cm^2}$) for three different droplet sizes, N=5000, 10000, 15000 from left to right. The upper panels (a-c) show the energy absorption in the He shell and in the Xe region  (as indicated) as a function of pulse delay. In the bottom panels (d-f) average charge states for inner and outer ionization as well as final average charge states including recombination are given for different time delays.}
\end{figure}

The simulations show that, irrespective of the nanodroplet size, charging begins in the pump pulse with tunnel ionization of the dopant atoms residing at the droplet center. Laser heating of the first released electrons induces an impact ionization avalanche that quickly increases the charge states of the dopant core and the closest He shells around it. Subsequently, triggered by the plasma produced in the core region, the rest of the He nanodroplet becomes ionized from inside-out and expands. Fig.~\ref{fig:md_xehe} shows the calculated delay dependence of key observables for three different droplet sizes, N$\,=\,$5000, 10000, 15000, from left to right.

%%%%%%%%%%%%%%%%%%%%%%%%%%%%%%%%%%%%%%%%%%%%%%%%%%%%%%%%%%%%%%%%%%%%%%%%%%%%%%%
%%%%%%%%%%%%%%%%%%%%%%%%%%%%%%%%%%%%%%%%%%%%%%%%%%%%%%%%%%%%%%%%%%%%%%%%%%%%%%%
Focusing on the smallest nanodroplet first, the following general conclusions can be extracted (see left panels in Fig.~\ref{fig:md_xehe}). The average inner charge state of He, i.\,e. the average number of electrons removed from He atoms, is significantly enhanced for short pulse delays. This reflects the density dependence of the inner ionization rates, as both the probability for electron-ion collisions as well as the influence of local plasma fields decrease with decreasing density. Note that no direct signature of resonant heating can be found in the pump-probe traces of inner ionization. In contrast to that, pronounced maxima occur in the energy absorption of the He nanodroplet (around 175\,fs) and the absorption in the Xe region (around 450\,fs), see Fig.~\ref{fig:md_xehe}(a). The number of continuum electrons (outer ionization) exhibits a similar dynamics as the total energy absorption, purple curve in Fig.~\ref{fig:md_xehe}(d). The maxima in absorption and outer ionization clearly indicates resonant heating of the nanoplasma within the droplet.

For a realistic comparison of theory and experiment, charge spectra have been determined by taking recombination processes into account. The orange line in Fig.~\ref{fig:md_xehe}(d) indicates the calculated final average charge states of the He atoms after applying our simplified recombination scheme. The effect of recombination is reflected by the difference between average inner ionization and final charge states. In contrast to the inner charge state, the final charge state shows a strong delay dependence over the whole investigated range including a pronounced maximum for a delay of 150\,fs and a shoulder near $\Delta t=$450\,fs. Obviously, the dynamics in the final charge states is a result of the interplay of inner ionization and recombination.

Because of the high temperature dependence of three-body-recombination, recombination is substantially suppressed in the case of enhanced heating of the nanoplasma. Comparison of inner and final charge states with the absorption in Fig.~\ref{fig:md_xehe}(a) shows that recombination is ineffective for delays with high absorption. This trend is most pronounced for resonant heating of the He shell. Furthermore, also the resonant heating of the Xe core is sufficient to notably suppress recombination and leads to the shoulder in the final charge around 450\,ps delay.
The peak-structure in the final charge states can be traced back to the combined action of inner ionization and resonant heating, in agreement with previous results~\cite{PeltzEPJD2011}.

A comparison of the results for the different droplet sizes reveals the same trends and shows that the above picture is generic for the pump-probe excitation of the doped nanodroplets and is in reasonable agreement with the experimental findings. For all sizes of the He matrix, a pronounced peak in the final charge state is predicted. The peak values appear for delay times similar to the experiment and are shifted towards longer delays with increasing size of the He droplet. The increase of the optimal delays can be traced back to the fact that the larger matrix requires more time to expand to resonant conditions.

\section{Conclusions}
We have studied the evolution of dopant-induced He nanoplasmas with a combination of experiment and classical MD simulations. High yields of He$^+$ and He$^{2+}$ ions are experimentally observed when doping He nanodroplets with a few Kr or Xe atoms upon irradiation with pairs of few-cycle IR laser pulses. The He ion signals exhibit a pronounced resonance feature as a function of the pump-probe delay time, which most sensitively depends on the average size of the He droplets. A linear increase in the optimal pump-probe delay as a function of the size of the nanodroplets clearly elucidates the dynamical role played by ionized He shells in the resonance of the composite nanoplasma. Our detailed numerical studies lead to the development of the following dynamical picture of the two-component nanoplasma evolution: (i) The whole interaction dynamics of the doped He nanodroplets is launched by tunnel ionization of the dopant atoms. Heating of the first released electrons induces an impact ionization avalanche that charges up the whole droplet from inside to outside. (ii) The observed delay dependence of the charge states can be clearly attributed to the varying recombination efficiency, which is minimal for resonant heating conditions. (iii) The optimal delay for high ion yields depends on the droplet size and the laser intensity (see Ref.~\cite{DoepEPJD2005}). Therefore the experimental data are the result of a convolution of the droplet size distribution and the intensity profile of the laser focus. This leads to broader structures and masks details such as the numerically observed sideband due to the nanoplasma resonance at the core. Nevertheless, the salient features, i.\,e. the shift and the broadening of the observed resonance structures as a function of droplet size, are reproduced in the simulations.

The present studies motivate further experimental and theoretical investigations of the dynamics in two-component nanoclusters. While the present study brings to light the active role of the He shells in the nanoplasma expansion dynamics, it also opens up interesting questions for further investigation such as the effect of the location of the dopant atoms within the droplet (center versus surface) on the nanoplasma ionization and expansion dynamics.

\subsection{Acknowledgments}
Support by DFG is gratefully acknowledged.
C.P. and T.F. gratefully acknowledge financial support from the DFG via SFB 652/2 and computer time provided by the High-Performance-Computing-Center for North Germany (HLRN).
M.K. acknowledges financial support through the ``Partner groups in India'' scheme of the Max Planck Gesellschaft.

\section*{References}
\bibliographystyle{nature}
%\bibliography{KrishnanEtAl_NJP2012}

\end{document}